# Gamma-rays Associated with Nearby Thunderstorms at Ground Level

Rebecca Ringuette[*], Michael L. Cherry, Douglas Granger, T. Gregory Guzik, Michael Stewart, John P. Wefel

Dept. of Physics & Astronomy, Louisiana State University, Baton Rouge, LA 70803, USA

**ABSTRACT:** The TGF and Energetic Thunderstorm Rooftop Array (TETRA) is an array of NaI scintillators located at rooftop level on the campus of Louisiana State University in Baton Rouge, Louisiana. From July 2010 through March 2014, TETRA has detected 28 millisecond-duration bursts of gamma-rays at energies 50 keV - 2 MeV associated with nearby (< 8 km) thunderstorms. The ability to observe ground-level Terrestrial Gamma Flashes from close to the source allows a unique analysis of the storm cells producing these events. The results of the initial analysis will be presented.

**INTRODUCTION**

Terrestrial Gamma-ray Flashes (TGFs), initially observed by the Burst And Transient Source Experiment (BATSE) in 1994, are millisecond bursts of gamma-rays produced by electrons accelerated upwards to energies of tens of MeV or more [Fishman et al., 1994; Grefenstette et al., 2009; Briggs et al., 2010; Tavani et al., 2011]. Events detected from satellite altitudes have been correlated with high resolution lightning data, generally positive polarity intracloud lightning [Hazelton et al., 2009]. (Positive polarity is needed to produce the upward beam of electrons and secondary photons necessary for detection of TGFs from space [Dwyer, 2003; Cohen et al., 2010].) Monte Carlo simulations have suggested that the relativistic runaway electron avalanche (RREA) process can produce these events at altitudes near thunderstorm tops [Dwyer and Smith, 2005]. This process generates ~$10^{17}$ runaway electrons as observed by the BATSE, Reuven Ramaty High Energy Solar Spectroscopic Imager (RHESSI), and Fermi Gamma-ray Burst Monitor (GBM) instruments [Briggs et al., 2010].

Most ground-level observation projects currently focus on correlating satellite-observed TGFs with lightning and measuring possible associated magnetic signatures [Cummer et al., 2011; Lu et al., 2011]. The International Center for Lightning Research and Testing (ICLRT) project, however, has reported two gamma-ray bursts, one in association with triggered lightning of negative polarity [Dwyer et al., 2004] and another in association with nearby negative polarity cloud-to-ground lightning [Dwyer et al., 2012]. TGFs associated with negative polarity lightning strikes, as with these ICLRT events, produce downward beams of photons which can be detected from the ground. ICLRT operates in a triggered mode, requiring either a triggered lightning current above 6 kA or the simultaneous trigger of two optical sensors.

Observations of TGFs from the ground are necessary to resolve several issues. Although hundreds of TGFs have been detected from space, satellites have difficulty distinguishing low flux events from background. Also, the location uncertainty of the majority of these events is 300 km or more – much larger than individual thunderstorm cells [Briggs et al., 2013]. With a ground-based array, we are able to detect nearby TGFs, associate them with specific portions of thunderstorms, and look for trends between the TGFs and the properties of the storms producing them. Here we present observations from July 2010 through March 2014 of twenty-eight TGF-like events in which 50 keV - 2 MeV gamma-rays are observed at ground level in shorter than 5 msec bursts associated with nearby lightning, typically of negative polarity. These observations increase the number of TGFs detected from the ground by a factor of 15, compared to the two previously known TGFs detected from the ground by ICLRT.

[*] Contact information: Rebecca Ringuette, Dept. of Physics & Astronomy, Louisiana State University, Baton Rouge, LA 70803, Rebecca.Ringuette@gmail.com

**GROUND-BASED TGF DETECTION ARRAY**

The TGF and Energetic Thunderstorm Rooftop Array (TETRA, described in more detail in Ringuette et al. [2013]) consists of an array of twelve 19 cm × 19 cm × 5 mm thallium-doped sodium iodide (NaI) scintillators designed to detect the gamma-ray emissions from nearby lightning flashes over the range 50 keV - 2 MeV. The scintillators are mounted in four detector boxes, each containing three sodium iodide detectors viewed by individual photomultiplier tubes (PMTs). The boxes are spaced at the corners of a ~700 × 1300 m$^2$ area on four high rooftops at the Baton Rouge campus of Louisiana State University at latitude 30.41º and longitude -91.18º. Unlike ICLRT, TETRA operates in a self-triggered mode, allowing for events to be recorded without requiring the direct detection of lightning.

Each TETRA detector box contains three NaI scintillator plates oriented at 30° from the zenith direction and separated by 120° in azimuth. Each NaI crystal is hermetically sealed between a 6.4 mm thick glass optical window on one flat face and a 0.75 mm thick Aluminum entrance window on the other face. An ultraviolet transmitting lucite lightguide is coupled to the glass window, and the light is viewed by an Electron Tubes 9390KB 130 mm PMT with a standard bialkali photocathode. The scintillator-PMT assemblies are housed in ~ 1″ thick plastic foam insulation to prevent rapid temperature changes.

The ADC-to-energy conversion is calibrated with radioactive sources ($^{22}$Na, $^{137}$Cs, $^{60}$Co). Individual detector energy resolution ranges from 9 to 13.5% full width half maximum at 662 keV and from 5.5 to 10.8% at 1.3 MeV. The total interaction probability in the NaI scintillators is 95% at 100 keV, 82% at 500 keV, and 10% at 1 MeV (with photoelectric interaction probabilities 93%, 26%, and 0.63% respectively). Beginning in October 2012, all boxes contain a bare PMT (photomultiplier tube without a scintillator) to check for electronic noise.

Data are accumulated for a day at a time for each of the four detector boxes individually. The daily analysis software selects events with signals corresponding to at least 50 keV deposited energy within 1 μsec. The data are then binned into 2 msec bins and assigned a timestamp. TETRA triggers are selected with counts/2 msec at least 20 standard deviations above the mean for the day. (For a typical average counting rate of 8900 min$^{-1}$ in a detector box above 50 keV, a 20 σ excess corresponds to 10 counts in the three PMTs in a detector box within a 2 msec window.) Once days with excessive electronic noise or other instrumental problems are removed, there are 909.16 days of live time and 2091 TETRA triggers.

**TETRA RESULTS**

TETRA triggers are compared to lightning data provided by the US Precision Lightning Network (USPLN) Unidata Program. Triggers occurring within seven seconds of lightning strikes reported within 5 miles (8 km) of the array are considered event candidates (ECs). From July 2010 through March 2014,

TETRA has recorded a total of twenty-eight event candidates.

The accidental rate of triggers coincident within 7 sec of a lightning flash that is less than 5 miles distant (i.e., events masquerading as ECs) is calculated based on the rate of TETRA triggers (due mainly to cosmic ray showers), the live time, and the duration of storm activity. The storm activity time is taken to be the sum of all time windows where there was lightning within 5 miles and 7 seconds and there was no electronic noise or other instrumental problems. For a total storm time of 21.09 hours, we calculate the expected number of ECs due to accidental triggers to be 2.02, assuming 100% lightning detection efficiency. The USPLN is part of the National Lightning Detection Network (NLDN), which has an efficiency above 99% in our area for cloud-to-ground lightning.

Figure 1 shows the time history of events measured in a single box during a 100 msec window centered on one of the event candidates. Background in the NaI detectors is mostly due to cosmic ray events. No events are observed in the bare PMT at the time of the trigger.

Of the 28 ECs, 5 occurred at times when accurate timing was not available due to network problems. Figure 2 compares data acquired within 7 seconds of lightning to the remaining data for the 23 ECs with accurate timing information. The distribution of events vs $\sigma$ within 7 seconds of a USPLN lightning strike within 5 miles is shown in black. The significance distribution of the remaining data has been normalized to the total storm activity time of the lightning distribution for comparison, shown in grey. The excess of events above 20 sigma in the lightning distribution (black) as compared to the normalized distribution (grey) indicates the association of the gamma-ray events with nearby lightning. (Note that three events involve seven separate coincident triggers between multiple detector boxes, so that there are 27 individual triggers shown in Figure 2 compared to the 23 ECs with accurate timing information, as discussed later.) A Kolmogorov-Smirnov test of the two distributions results in a D parameter of 0.87 compared to the value of 0.25 reported by Ringuette et al. [2013] based on the initial analysis of the events though 2012. This distinguishes the ECs from the background distribution with high confidence.

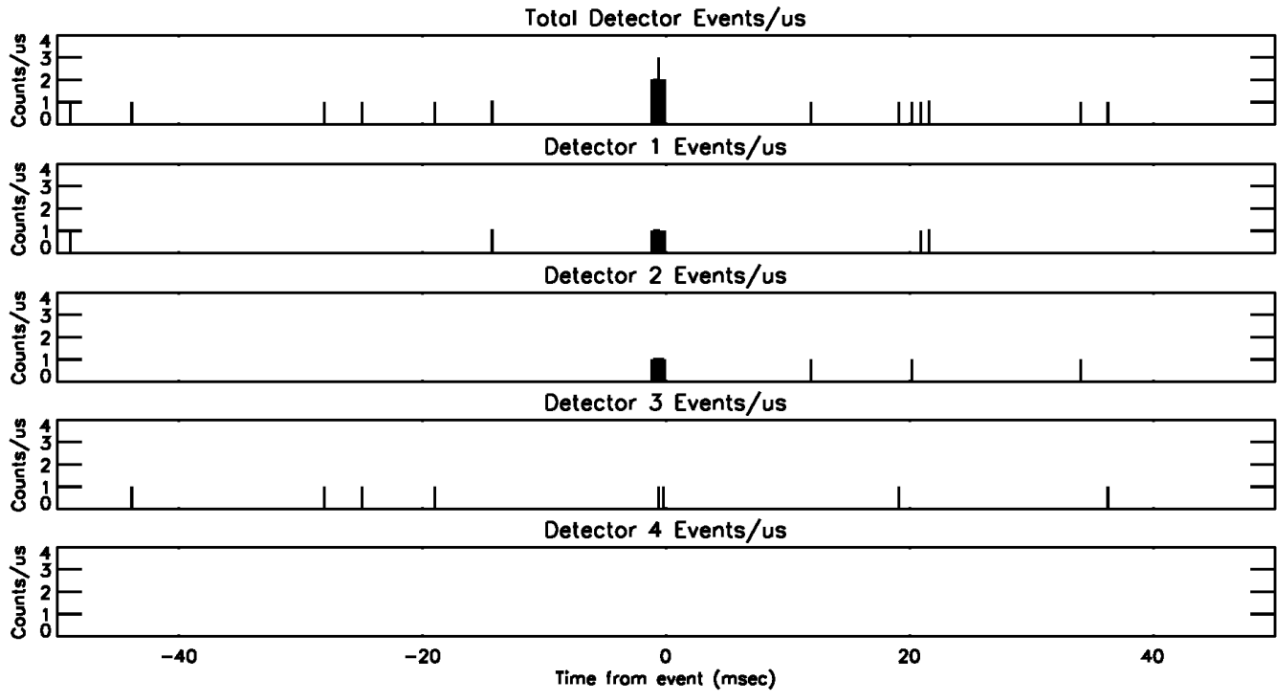

Figure 1: Individual PMT hits shown in 100 msec window centered on event candidate. Upper frame shows total events in two NaI PMTs in a single detector box; next two frames show individual NaI PMT time histories; third frame shows events in a high resolution $LaBr_3$ scintillator included in this detector box (Ringuette et al. [2013]); lowest frame shows absence of events in bare PMT.

The properties of the twenty-eight events observed by TETRA are reported in Table 1. Events detected in more than one detection box within the timing uncertainty are classified as Coincident Event Candidates (CECs) and are listed in the top portion of the table. The remaining events detected in the 2011 and 2012 seasons are listed in the second and third sections of the table, with the 2013 events at the bottom of the table. In this table each event date, trigger time and uncertainty is listed, along with the number of lightning flashes detected within ±2.5 minutes and 5 miles and the cloud density above TETRA. Also listed for each EC is the distance to the lightning stroke closest in time to the event trigger, the current, the number of gamma-rays detected in the EC, and the $T_{90}$ duration of the event. ($T_{90}$, the time over which a burst emits from 5% to 95% of its total measured counts in a single detector box, was calculated by considering all events detected within a ±3 msec window around the trigger time, discarding the first and last 5% of timestamps for each event, and recording the time difference between the first and last events remaining. The uncertainty in the T$_{90}$ determination is approximately ±200 μsec based on a Monte Carlo simulation.) The number of sigma above the mean and the expected number of CECs due to random triggers are listed in the last two columns for each event.

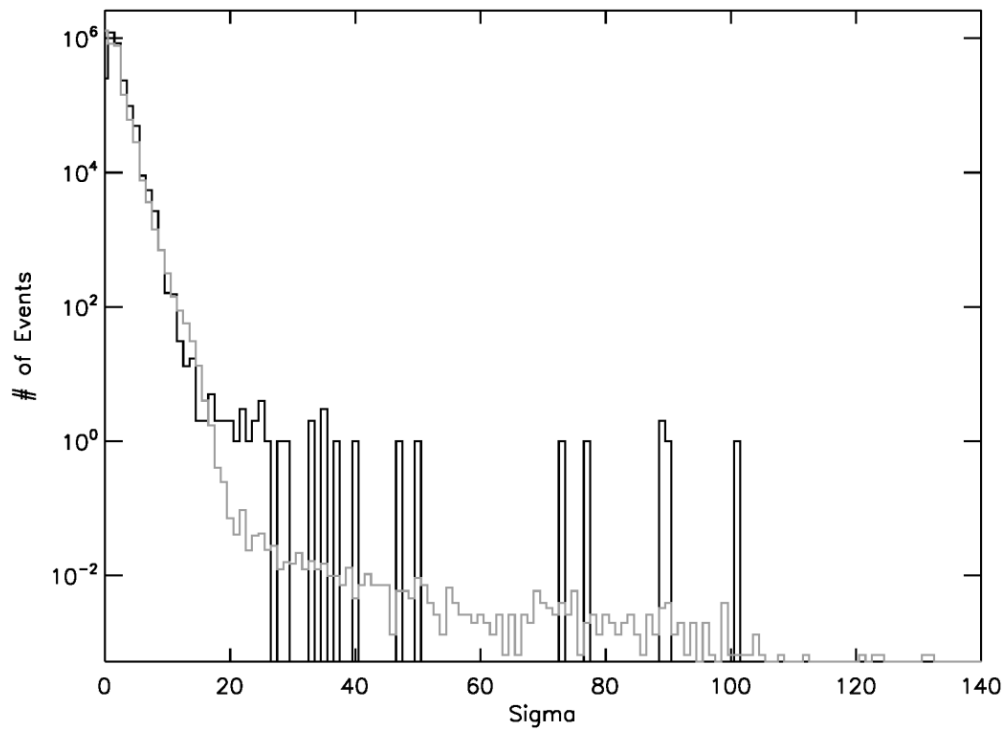

Figure 2: Distribution of events with significance σ for July 1, 2010 through October 31, 2013. Distribution of events within 7 seconds of nearby (< 5 miles) lightning is shown in black. Distribution of all data, normalized to 0.88 days of live time, is shown in grey, showing excess of lightning-associated ECs at σ > 20.

| Date | Trigger Time (CST) (hh:mm:ss) | Max Lightning Rate within 5mi. (sec$^{-1}$) | Cloud Density (dBZ) | Storm Type | # Flashes within 5mi. and 5min. | Lightning Distance (mi) | Lightning Current (kA) | T$_{90}$ Event Duration (μs) | Total γ Rays Detected | Total Energy (MeV) | σ Above Mean | Probability of CEC |
|---|---|---|---|---|---|---|---|---|---|---|---|---|
| 31 Jul 2011 | 16:21:44.976 | 2 | 45 | Coastal | 12 | 1.4 | -43.6 | 702 | 19 | 8.8 | 25.1 | 1.7E-06 |
| 31 Jul 2011 | 16:21:45.300 | 2 | 45 | Coastal | 12 | 1.8 | -29.1 | 1326 | 23 | 9.6 | 25.1 | 1.7E-06 |
| 18 Aug 2011 | 17:57:38.986 | 4 | 50 | Coastal | 40 | 1.3 | -23.4 | 1318 | 38 | 16.5 | 22.4 | 1.2E-13 |
| 24 Feb 2011 | 23:11:15.787 | 3 | 45 | Front | 1 | 2.9 | -20.9 | 953 | 20 | 1.7 | 24.6 | - |
| 29 Jul 2011 | 10:38:58.932 | 6 | 45 | Coastal | 42 | 0.4 | -57.7 | 153 | 8 | 4.8 | 22.6 | - |
| 18 Aug 2011 | 17:57:39.202 | 4 | 50 | Coastal | 40 | 1.3 | -23.4 | 24 | 7 | 3.6 | 23.8 | - |
| 12 Mar 2012 | 11:30:16.500 | 6 | 45 | Front | 4 | 1.6 | -81.3 | 1997 | 7 | 3.2 | 20.7 | - |
| 2 Apr 2012 | 12:29:30.554 | 3 | 50 | Coastal | 8 | 0.6 | -29.9 | 464 | 21 | 15.8 | 90.0 | - |
| 4 Apr 2012 | 02:49:21.900 | 5 | 55 | Front | 21 | 1.9 | -158.4 | 515 | 24 | 21.3 | 77.2 | - |
| 5 Aug 2012 | 14:43:35.661 | 7 | 40 | Coastal | 16 | 0.6 | -56.5 | 392 | 11 | 6.1 | 35.3 | - |
| 6 Aug 2012 | 19:17:33.359 | 5 | 50 | Coastal | 1 | 0.8 | -23.1 | 465 | 10 | 4.5 | 21.8 | - |
| 9 Aug 2012 | 15:27:29.804 | 4 | 50 | Front | 21 | 0.4 | -27.8 | 2412 | 12 | 2.9 | 25.6 | - |
| 9 Aug 2012 | 15:28:36.070 | 4 | 50 | Front | 27 | 0.9 | -36.7 | 4217 | 24 | 7.4 | 35.7 | - |
| 9 Aug 2012 | 15:28:36.560 | 4 | 50 | Front | 27 | 0.8 | -19.2 | 146 | 12 | 8.0 | 29.7 | - |
| 6 Jun 2012 | 15:37:31 | 6 | 55 | Coastal | 40 | - | - | 759 | 36 | 17.5 | 85.3 | - |
| 6 Jun 2012 | 15:44:18 | 6 | 55 | Coastal | 16 | - | - | 609 | 14 | 8.5 | 46.9 | - |
| 6 Jun 2012 | 19:29:43 | 6 | 55 | Coastal | 33 | - | - | 2376 | 24 | 9.7 | 51.9 | - |
| 6 Jun 2012 | 19:31:21 | 6 | 55 | Coastal | 19 | - | - | 746 | 31 | 16.2 | 45.6 | - |
| 6 Jun 2012 | 19:36:41 | 6 | 55 | Coastal | 18 | - | - | 604 | 28 | 21.1 | 57.7 | - |
| 14 Apr 2013 | 01:26:02.390 | 4 | 45 | Coastal | 2 | 0.7 | -46.9 | 1552 | 9 | 3.9 | 51.4 | - |
| 24 Apr 2013 | 07:11:37.894 | 5 | 50 | Front | 24 | 1.9 | -64.8 | 616 | 7 | 1.6 | 25.6 | - |
| 10 May 2013 | 03:51:57.412 | 5 | 55 | Front | 166 | 1.3 | -23.9 | 1032 | 29 | - | 101.3 | - |
| 10 May 2013 | 03:51:58.116 | 5 | 55 | Front | 163 | 1.3 | -23.9 | 80 | 6 | 2.1 | 25.2 | - |
| 22 Jun 2013 | 14:31:28.794 | 5 | 50 | Coastal | 7 | 1.7 | -33.8 | 159 | 8 | 1.9 | 25.8 | - |
| 22 Jun 2013 | 14:52:49.063 | 5 | 50 | Coastal | 6 | 1.3 | -48.9 | 1757 | 15 | 5.6 | 50.4 | - |
| 29 Jun 2013 | 04:24:11.550 | 4 | 40 | Front | 17 | 1.7 | -32.9 | 732 | 14 | 7.0 | 89.7 | - |
| 29 Jun 2013 | 04:24:11.614 | 4 | 40 | Front | 17 | 1.7 | -32.9 | 164 | 4 | 3.0 | 31.6 | - |
| 13 Sep 2013 | 18:11:13.263 | 5 | 50 | Coastal | 39 | 1.4 | -35.3 | 1539 | 18 | 6.2 | 40.9 | - |

Table 1: Properties of the 28 Event Candidates. CECs are listed in the top section; ECs for which the absolute timing uncertainty is known are listed in the second section; and ECs for which the absolute timing uncertainty is unknown are listed in the third section of the table. The date and time of each EC trigger are listed (Columns 1-2), along with the properties of the storm associated with each event (Columns 3-6). The properties of the associated lightning (Columns 7-8), event duration (Column 9), number of gamma-rays detected (Column 10), total energy (Column 11), event significance (Column 12), and random probability of CECs (Column 13) are also listed for each event.

For the original data acquisition software, the absolute timing was checked up to 600 times per 24-hour period based on a comparison of the onboard system clock and network timing. The magnitudes of the resulting corrections were on the order of a few milliseconds. Five ECs were detected on 6 June 2012 during a period when accurate trigger-lightning time differences were not recorded due to network timing difficulties. These ECs were correlated with two intense thunderstorms that passed directly over TETRA. The timing software was upgraded for the 2013 season to include a Global Positioning System (GPS) disciplined clock. This GPS timing software performs time corrections every second with correction magnitudes of tens to hundreds of nanoseconds. Consequently, the timing uncertainties associated with the 2013 events are less than 60 nanoseconds while the timing uncertainties of earlier events are 0.5 to 5 milliseconds. (Two ECs were detected on Aug 5 and 6, 2012 during the testing stage for this software; the exact timing uncertainty for these events is unknown but expected to be within ±200 nsec.)

In each of the 28 events, 7 to 38 γ-rays were detected within a time window of less than 5 msec, with the total energy deposited per event ranging from 2 to 32 MeV. The distances to the nearest lightning flashes were 0.4 - 2.9 miles. For 23 events, absolute timing was available with ~5 msec accuracy or better. For each of these events, lightning was observed within 7 seconds of the trigger time. Nine of these events were associated with negative polarity cloud-to-ground lightning detected within 6 msec of the trigger, suggesting an association with negative polarity cloud-to-ground lightning, as with the ICLRT events.

The expected number of CECs due to random triggers is small. Given an initial EC with counting rate in one box in excess of 20 σ above the daily average, the likelihood that a second or third trigger occurred at random in another box within a timing uncertainty of 2 msec on the same day is estimated as $(2 \times 2 \text{ msec} \times N/86400 \text{ sec})^{b-1}$, where N is the total number of random 20 σ triggers detected per day and b is the number of boxes triggered in the event. Multiplying by the number of ECs then gives the expected number of spurious CECs involving two boxes occurring by chance as $1.7 \times 10^{-6}$, as listed in Table 1. The CEC on 18 Aug 2011 was detected on three boxes within the timing uncertainties, resulting in a random probability of $1.2 \times 10^{-13}$. Although three CECs were reported in the 2011 and 2012 seasons, no CECs were detected in the 2013 season.

The dark solid histogram in Figure 3 shows the deposited energy spectrum of the 28 Event Candidates, with events observed up to 2.2 MeV. It should be emphasized that, with TETRA's thin detectors, only a portion of the incident gamma-ray energy is actually detected. Between 200 keV and 1.2 MeV, the EC spectrum is fit with a power law $E^{-\alpha}$, with $\alpha = 0.90 \pm 0.14$ and $\chi^2$/degree of freedom = 0.95 (dark dashed line). On the same figure, the grey line shows the spectrum of non-EC triggers (i.e., triggers not associated with lightning within 5 miles and 7 seconds); this spectrum is softer, with a best fit power law index $\alpha = 1.58 \pm 0.04$ and $\chi^2$/degree of freedom = 2.0 (grey dashed line).

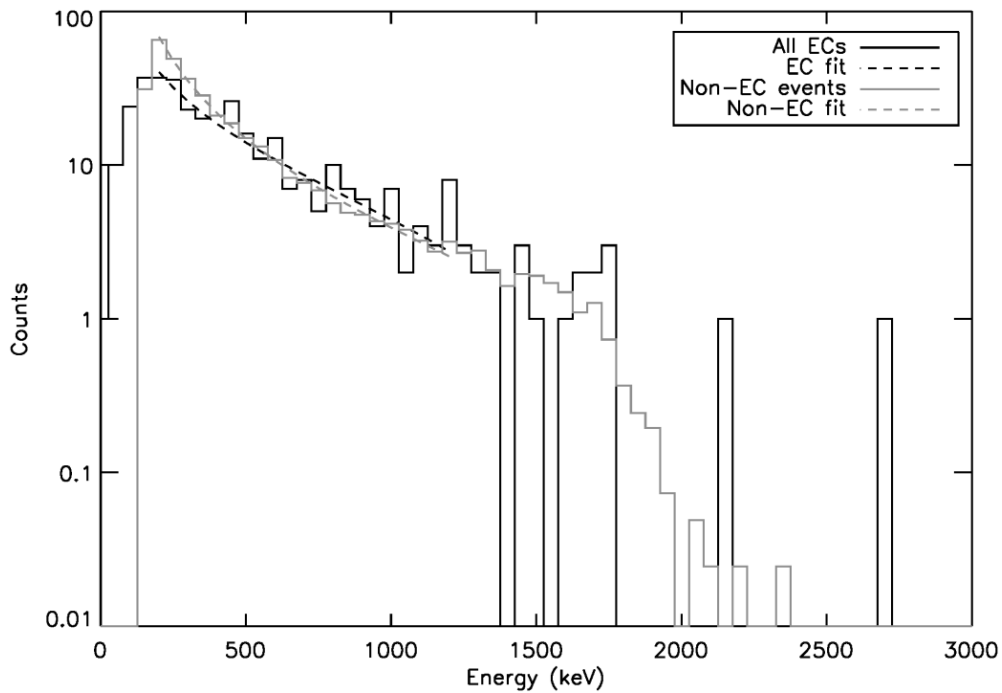

Figure 3: Spectra of Event Candidates and non-EC TETRA triggers. Spectrum of ECs is shown in black. Spectrum of non-EC triggers (triggers not associated with lightning nearby in time and distance) is shown in grey. Power law fits between 200 keV and 1200 keV are shown with dotted lines.

The properties of the nine events detected in 2013 are similar to the previously reported events [Ringuette et al., 2013] with the exception of lightning associations. Nine of the 19 previously reported events were associated with lightning. However, none of the 2013 events were linked to lightning within 100 milliseconds of the event time and 5 miles (8 km) of TETRA. This suggests that some gamma-ray events observed from the ground are not directly associated with lightning [Connaughton et al., 2013] or are produced by intracloud strikes which are not easily detected by USPLN [Strader et al., 2013].

The twenty-eight events presented here were detected with TETRA from July 2010 to March 2014 with nine events in 2013 and none seen between October 2013 and March 2014. The majority of these events occurred from June to August, when storms in southern Louisiana tend to be associated with disturbances in the Gulf of Mexico rather than frontal lines. However, almost half of the events in Table 1 were associated with fronts, hinting that the source of the storm associated with the TGF may not be as important as the strength and maturity of the storm itself. The storms producing these events are discussed in the next section.

**STORM ANALYSIS**

Although efforts have been made to correlate TGF production with storm evolution based on lightning flash rates [Smith et al., 2010], there has been only one study to date on the radar properties of the storms that produce TGFs [Splitt et al., 2010]. Splitt et al. performed a population study of storms associated with TGFs detected by RHESSI, but was unable to analyze the maturity stage of the storms due to lack of detailed radar information and the high uncertainty in the location accuracies of the events. With TETRA's reliable detection of TGFs from the ground, the detailed characteristics of the associated storms can be analyzed.

The thunderstorms associated with TETRA events fall into two general categories: single cell thunderstorms and squall lines. Single cell thunderstorms are generally short-lived (a few hours) and only rarely produce severe weather. In Louisiana, these storms are produced by warm, moist updrafts common to coastal environments and can approach from all directions. Squall line thunderstorms form the basis of frontal lines but also occur with some summer storms. In Louisiana, frontal lines usually approach from the west while summer thunderstorms associated with squall lines approach from the northwest, north and northeast. Of the 17 storms producing the 28 events, eight were single cell thunderstorms and nine were squall lines.

Half of the TGFs observed by TETRA were produced by variations of single cell thunderstorms. An example of a single cell thunderstorm at the time of a TETRA TGF is shown in Figure 4. The smoothed radar scan is shown, produced by the GR2Analyst software using radar data acquired from the National Oceanic and Atmospheric Administration (NOAA) National Climatic Data Center website. The front of the storm is located near White Castle, LA and is moving SE towards Donaldsonville, LA. The colors in the image correspond to the varying densities of the cloud at the lowest elevation (see scale in image). As is common for thunderstorms, the main updraft of this storm is located towards the front of the storm, shown by the purple and red areas in the southern portion of Figure 4. The downdraft is located behind this area and fans out to the northwest, north and northeast. The green triangle in the southern section of the image indicates the location of hail less than one inch in diameter detected by Next Generation Radar (NEXRAD) [Stumpf et al., 1997].

Figure 5 shows a three-dimensional image of this thunderstorm. The storm is viewed from the east from slightly above the image plane. The smoothed radar image is shown on the bottom layer of the volume for reference. Altitude is indicated by the horizontal grey lines in tens of thousands of feet. The white arrow indicates the movement of the storm. Iso-density surfaces of 30, 40, and 50 dBZ are shown in green, yellow, and red respectively, to visualize the structure of the thunderstorm cloud. As expected, the updraft shown by the dense clouds (on the left of the image) is also correlated with the tallest clouds of the storm. The trailing downdraft is located towards the rear of the storm (right side of the image) and is associated with clouds of decreasing altitude as distance from the updraft increases. This storm occurred during Louisiana's winter season, producing clouds with a maximum altitude of 10.9 km (35.9 kilofeet). For comparison, summer thunderstorms in Louisiana often reach up to 15-18 km (50 to 60 kft), altitudes comparable with TGF production heights [Dwyer and Smith, 2005; Grefenstette et al., 2008; Shao

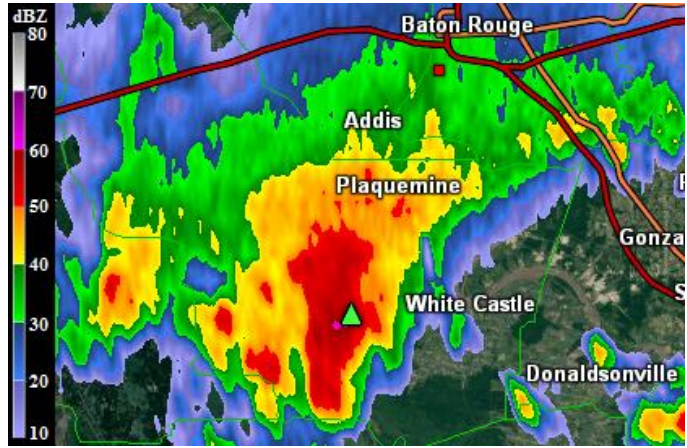

Figure 4: Smoothed radar image of the single cell thunderstorm producing the TETRA TGF on 12 Mar 2012. Colors in the image correlate to the density of the cloud in decibels at the lowest elevation angle as indicated in the scale at left. The green triangle indicates the location of hail. The red square shows the location of TETRA. Local interstates and highways are shown with red and orange lines near the top of the image. The locations of various cities are also labeled for reference. The image is approximately 90 km by 50 km.

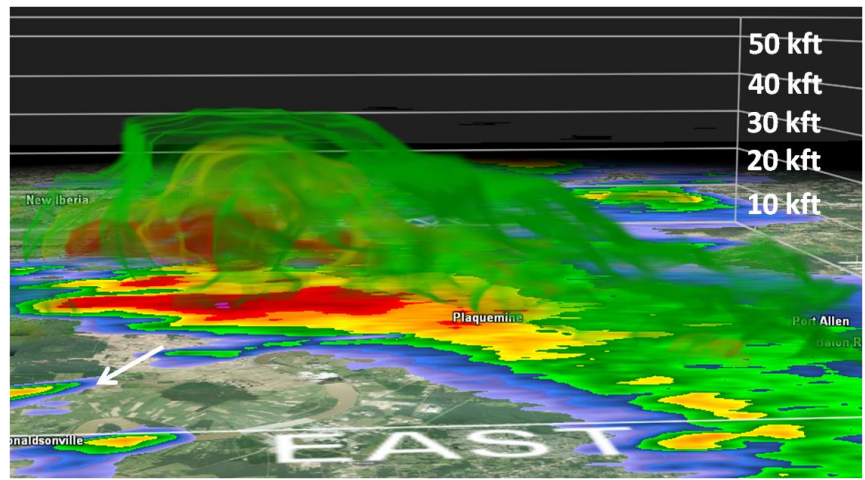

Figure 5: Three dimensional radar image of the single cell thunderstorm producing the TETRA TGF on 12 Mar 2012. Altitude is indicated by the grey horizontal lines in tens of thousands of feet. The smoothed radar image in Figure 4 is shown on the bottom plane for comparison. Iso-density surfaces of 30 dBZ, 40 dBZ and 50 dBZ are shown in green, yellow and red, respectively. The white arrow indicates the direction of the thunderstorm movement. The bottom plane of the image is approximately 90 km by 50 km.

et al., 2010; Gjesteland et al., 2010; Cummer et al., 2011; Xu et al., 2012].

Storms associated with squall lines produced the remaining half of the observed TGFs. As an example, Figure 6 shows the squall line thunderstorm that produced the TGF on 29 Jun 2013 at the time of the image. The main squall line extends from Livingston, LA to Breaux Bridge, LA, approximately 120 km in length. The trailing clouds extend north to New Roads, LA (approximately 50 km from the front of the storm). The storm is moving southeast toward the Gulf of Mexico. As in Figure 4, the colors in the image correlate to the varying densities of the cloud at the lowest elevation angle. Green triangles in the southern section of the image indicate the location of hail less than an inch in diameter detected by NEXRAD.

For the 17 storms observed to produce TGFs, the position of the array relative to the updraft and the maturity of the storms at the time of the events were recorded. For 14 storms, events were associated with collapsing cloud formations.

One such storm on 22 Jun 2013 produced two TGFs within ~ 20 minutes. Figure 7 shows the smoothed radar image of the storm at the time of the first event. The storm is located above TETRA and is gradually moving southwest toward Addis, LA. The main updraft of the single-cell storm is directly above TETRA when the TGFs are observed.

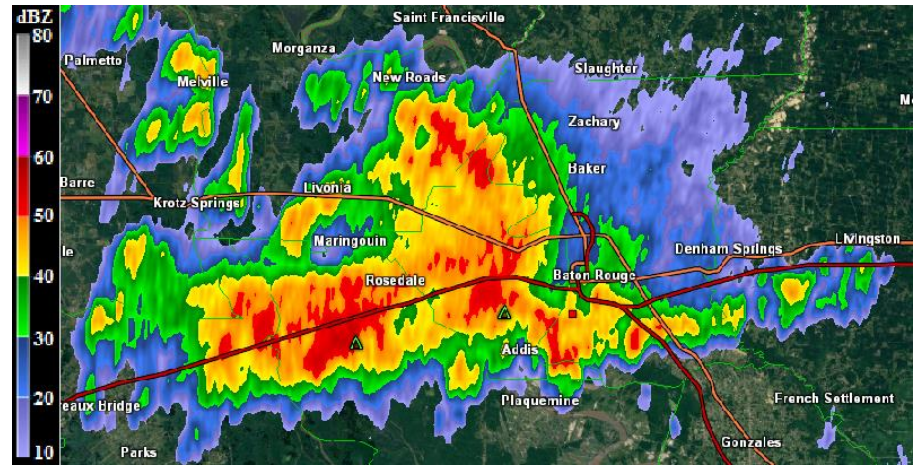

Figure 6: Smoothed radar image of a squall line of thunderstorms producing the TETRA TGF on 29 Jun 2013. Details similar to Figure 4. The image is approximately 120 km by 60 km.

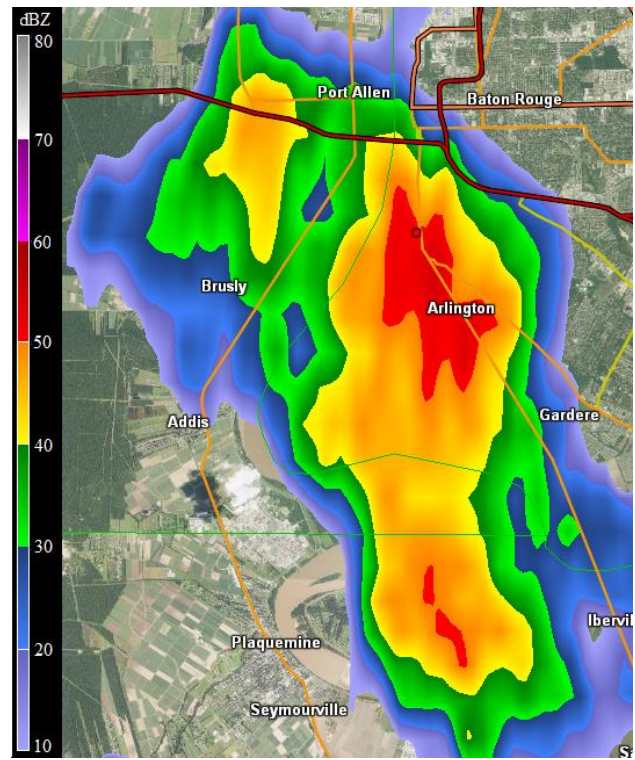

Figure 7: Smoothed radar image of the storm producing two TETRA TGFs on 22 Jun 2013. Image was taken at the time of the first TETRA TGF (20:33 UTC). The image is approximately 25 km by 40 km.

Figure 8 shows a time sequence of radar images of the storm taken every 4-5 minutes (as soon as the previous scan is completed), showing the behavior of the storm over 45 minutes. The colored cloud surfaces in the image again correspond to iso-density surfaces of 30 (green), 40 (yellow) and 50 (red) decibels. The storm is viewed from the northeast from a slightly tilted angle and the altitudes are given in tens of thousands of feet. The smoothed radar image at each time is shown on the bottom plane. The time of the scan is given at the top of each image. The radar images taken closest to the time of the TGFs are labeled.

The sequence begins during the initial intensifying stage of the storm. In the first image, the updraft is beginning to form a cloud tower directly above TETRA. By the third image (20:33 UTC), the cloud tower has reached 13.90 km and subsequently collapses after the TGF. A second cloud tower forms by 20:47 UTC, reaching 13.81 km and also collapses after the second TGF. After these two events, the storm completely dissipates.

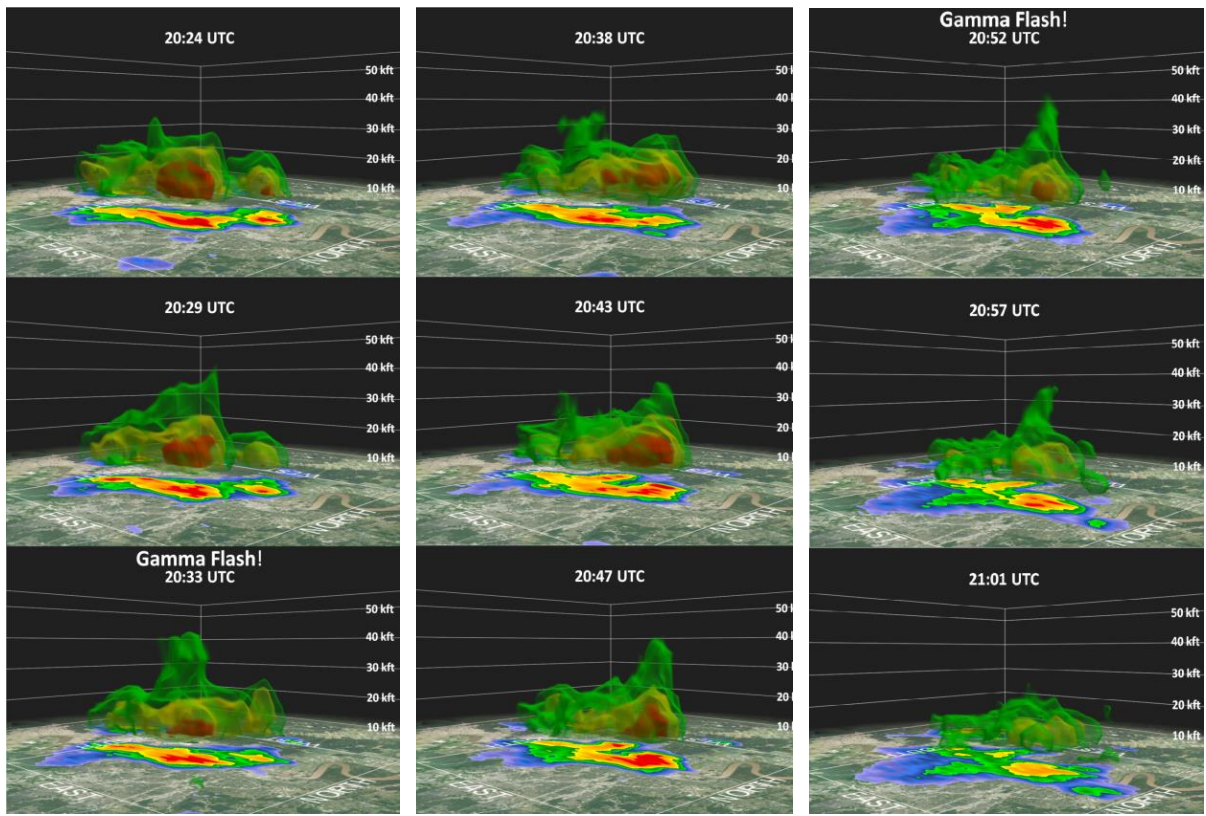

Figure 8: Three dimensional radar image sequence of the thunderstorm producing two TETRA TGFs on 22 Jun 2013. Time is indicated at the top of each image. The bottom plane of each image is approximately 30 km by 30 km.

**CONCLUSIONS**

Here we have presented data for 28 gamma-ray events observed from July 2010 to March 2014 with a self-triggered ground array, suitable for observing weak events from nearby distances without a bias caused by a lightning trigger. The event durations reported here are similar to the durations reported by satellites, although the energy range of the spectra and the low statistics prevent comparison with the TGF spectra observed from space. Ten of these gamma-ray events occur within 100 msec and 5 miles of lightning. The remaining gamma-ray events are either not correlated with nearby lightning [Connaughton et al., 2013], or the associated cloud-to-ground lightning strike was missed by the lightning network, or the event was due to intracloud lightning that was not detected by the lightning network [Strader et al., 2013].

An average of 20 photons were observed per gamma-ray event with an average energy of 500 keV per photon. Assuming isotropic emission at a distance of 1 mile, these events then require in excess of $\sim 10^{18}$ photons at the source [Ringuette et al., 2013] – several magnitudes higher than the 2009 ICLRT event [Dwyer et al., 2004] and a factor of ten higher than the brightest TGFs reported by BATSE, RHESSI, and the Gamma-ray Burst Monitor (GBM) [Briggs et al., 2010], suggesting that either the ground level TETRA events are beamed, or they are distinctly different from the ICLRT events.

The analysis presented here correlates 23 TGFs with collapsing cloud formations. Although the association of such a small number of TGFs with collapsing clouds does not classify TGFs as an indicator of decreasing storm strength, it does serve as an example for future work. In order to test this relationship, other TGF-producing storms should be analyzed in a similar fashion. If TGFs can indeed be established as a precursor of storm collapse, then TGFs may potentially be used in conjunction with other radar properties in the prediction processes of meteorologists.

**ACKNOWLEDGMENTS**
Lightning data were provided by the US Precision Lightning Network Unidata Program. The weather data used for the radar analysis were from NOAA NCDC. Radar analysis was performed using GR2 Analyst (v2.00). The authors would like to thank B. Ellison, D. Smith, S. Baldridge, and N. Cannady for their assistance with the project, and J. Fishman for valuable discussions. This project has been supported in part by NASA/Louisiana Board of Regents Cooperative Agreement NNX07AT62A and the Curry Foundation. R. Ringuette appreciates graduate fellowship support from the Louisiana Board of Regents.

**REFERENCES**
Boccippio, D. J., K. L. Cummins, H. J. Christian, and Goodman, S. J. (2001), Combined satellite- and surface-based estimation of the intracloud-cloud-to-ground lightning ratio over the continental United States. Mon. Wea. Rev. 129, 108.
Briggs, M. S., et al. (2010), First results on terrestrial gamma-ray flashes from the Fermi Gamma-ray Burst Monitor. J. Geophys. Res. 115, A07323.
Briggs, M.S., et al. (2013), Terrestrial Gamma-ray Flashes in the Fermi Era: Improved Observations and Analysis Methods. J. Geophys. Res. Space Physics 118.
Cohen, M. B., et al. (2010), A lightning discharge producing a beam of relativistic electrons into space. Geophys. Res. Lett. 37, L18806.
Connaughton, V., et al. (2013), Radio signals from electron beams in Terrestrial Gamma-ray Flashes. J. Geophys. Res. 118, 2313-2320.
Cummer, S. A., et al. (2011), The lightning-TGF relationship on microsecond timescales. Geophys. Res. Lett. 38, L14810.
Dwyer, J. R. (2003), A fundamental limit on electric fields in air. Geophys. Res. Lett. 30, 20, 2055.
Dwyer, J. R., et al. (2004), A ground level gamma-ray burst observed in associated with rocket-triggered lightning.

Geophys. Res. Lett. 31, L05119.

Dwyer, J. R. and D. M. Smith (2005), A comparison between Monte Carlo simulations of runaway breakdown and terrestrial gamma-ray flash observations. Geophys. Res. Lett. 32, L22804.

Dwyer, J. R., et al. (2012), Observation of a gamma-ray flash at ground level in association with a cloud-to-ground lightning return stroke. J. Geophys. Res. 117 A10303.

Fishman, G.J., et al. (1994), Discovery of intense gamma-ray flashes of atmospheric origin. Science 264, 5163, pp. 1313-1316.

Gjesteland, T., N. Østgaard, P. H. Connell, J. Stadsnes and G. J. Fishman (2010), Effects of dead time losses on terrestrial gamma-ray flash measurements with the Burst and Transient Source Experiment. J. Geophys. Res. 115, A00E21.

Grefenstette, B.W., D.M. Smith, J.R. Dwyer, and G.J. Fishman (2008), Time evolution of terrestrial gamma-ray flashes. Geophys. Res. Lett. 35, L06802.

Grefenstette, B. W., D. M. Smith, B. J. Hazelton and L. I. Lopez (2009), First RHESSI terrestrial gamma-ray flash catalog. J. Geophys. Res. 114, A02314.

Hazelton, B. J., et al. (2009), Spectral dependence of terrestrial gamma-ray flashes on source distance. Geophys. Res. Lett. 36, L01108.

Lu, G., et al. (2011), Characteristics of broadband lightning emissions associated with terrestrial gamma-ray flashes. J. Geophys. Res. 116, A03316.

Ringuette, R., et al, (2013), TETRA observation of gamma-rays at ground level associated with nearby thunderstorms. J. Geophys. Res. Space Physics 118, 7841.

Shao, X-M., T. Hamlin and D.M. Smith (2010), A closer examination of terrestrial gamma-ray flash-related lightning processes. J. Geophys. Res. 115, A00E30.

Splitt, M. E., S. M. Lazarus, D. Barnes, et al. (2010), Thunderstorm characteristics associated with RHESSI identified terrestrial gamma flashes. J. Geophys. Res. 11, A00E38.

Smith, D.M. et al. (2010), Terrestrial gamma-ray flashes correlated to storm phase and tropopause height. J. Geophys. Res. 115, A00E49.

Strader, S. M., W. S. Ashley, and G. D. Herbert (2013), A comparison and assessment of the USPLN and ENTLN. 93rd American Meteorology Society Annual Meeting, Austin, TX, Jan 9, 2013.

Stumpf, G. J., A. Witt, E. D. Mitchell, et al. (1998), The National Severe Storms Laboratory Mesocyclone Detection Algorithm for the WSR-88D. Weather and Forecasting 13, 304.

Tavani, M., et al. (2011), Terrestrial gamma-ray flashes as powerful particle accelerators. Phys. Rev. Lett. 106, 018501.

Xu, W., S. Celestin, and V. P. Pasko (2012), Source altitudes of terrestrial gamma-ray flashes produced by lightning leaders. Geophys. Res. Lett. 38, L08801.